\documentstyle[referee]{laa}

\begin{document}
\thesaurus{07  %A&A Section: Solar system
	   (07.09.1; % Interplanetary medium
	    07.13.1;  % Meteoroids
	   )}

\title{On Radiation Pressure and the Poynting-Robertson Effect
for Fluffy Dust Particles}
\author{J.~Kla\v{c}ka}
\institute{Astronomical Institute,
   Faculty for Mathematics, Physics and Informatics \\
   Comenius University,
   Mlynsk\'{a} dolina, 842~48 Bratislava, Slovak Republic; \\
   E-mail: klacka@fmph.uniba.sk}
\date{}
\maketitle

\begin{abstract}
Equation of motion for real dust particle under the action of electromagnetic
radiation is more general than equation of motion corresponding to standardly
used Poynting-Robertson effect (P-R effect). As a consequence, orbital
evolution of particles may significantly
differ from that corresponding to the P-R effect. The paper discusses recently
published (Icarus, June 2002) derivation of equation of motion, which
is in contradiction with known relativistically covariant formulation.
The ``new'' derivation does not respect fundamental physical
laws (law of conservation of energy, law of conservation of momentum) which
must hold in any frame of reference.
Application of the derived ``general'' equation of motion
to the special case treated by Einstein in 1905 yields result which is not
consistent with Einstein's result. Correct solution is presented.


\end{abstract}

\section{Introduction}
Relativistically covariant equation of motion for arbitrarily shaped (dust)
particle under the action of electromagnetic radiation was derived by
Kla\v{c}ka (2000a). A more simple derivations to the first order
in $v/c$ ($\vec{v}$ is velocity of the particle, $c$ is the speed of light)
were presented by Kla\v{c}ka (2000b), Kla\v{c}ka and Kocifaj (2001a), where
also application to orbital evolution of micron-sized meteoroids can be found.
Application of the general equation of motion to larger bodies (e. g.,
meteor-sized bodies, asteroids) can be found in Kla\v{c}ka (2000c). As for
review papers of various derivations (including relativistically covariant
forms) we refer to Kla\v{c}ka (2001) and Kla\v{c}ka (2002a).

Important property of the above discussed derivations and presentations of
general equation of motion is that it covers not only Poynting-Robertson
effect (P-R effect; Robertson 1937, Kla\v{c}ka 1992), but also Einstein's
example (Einstein 1905) as special cases -- as it is required
for more general theory. Equation of motion is expressed in terms of
particle's optical properties standardly used in optics for stationary
particles -- various optical properties can be taken into account.

Application of the general equation of motion for orbital evolution of
meteoroids was presented in Kla\v{c}ka and Kocifaj (2001a, 2001b, 2002),
Kocifaj and Kla\v{c}ka (2002a, 2002b).

In spite of easy accessible of at least some of the above referenced
papers a new paper by Kimura et al. (2002) was published, now. Kimura's paper
derives and presents a ``new'' derivation and completely ignores the above
referenced papers, most fundamental of which has been known to Kimura et al.
New paper in Icarus may generate an impression that the older presentations
are incorrect and Kimura's paper is right. In order to enable
orientation in physics of more general equation of motion than
that corresponding to the P-R effect, it is inevitable
to make comments on Kimura's presentation, corresponding to
presentations by Kla\v{c}ka and Kocifaj (1994) and Kocifaj et al. (2000);
as for some optical properties of dust particles, see also Kimura (2000).

\section{Equation of motion for arbitrarily shaped dust particle}
Relativistically covariant equation of motion may be expressed as
\begin{equation}\label{1}
\frac{d ~p^{\mu}}{d~ \tau} = \frac{w^{2}~S~A'}{c^{2}} ~
	 \sum_{j=1}^{3} ~Q_{j} ' ~ \left (
	 c ~ b_{j}^{\mu} ~-~ u^{\mu}  \right ) ~,
\end{equation}
where $p^{\mu}$ is four-vector of the particle of mass $m$
\begin{equation}\label{2}
p^{\mu} = m~ u^{\mu} ~,
\end{equation}
four-vector of the world-velocity of the particle is
\begin{equation}\label{3}
u^{\mu} = ( \gamma ~c, \gamma ~ \vec{v} ) ~,
~~\gamma = 1 / \sqrt{1 - \vec{v} ^{2} / c^{2}} ~,
\end{equation}
four-vectors
\begin{equation}\label{4}
b_{j}^{\mu} = ( 1 / w_{j} ) ~( 1 , \vec{e}_{j} ) ~, ~~ j = 1, 2, 3 ~,
\end{equation}
$\vec{e}_{1}$ is unit vector of the incident radiation,
the system of unit vectors $\{ \vec{e}_{j} ', j = 1, 2, 3 \}$ measured
in the rest frame of the particle forms an orthogonal basis,
S is flux density of the incident radiation energy, $A'$ is geometrical
cross section of a sphere of volume equal to the volume of the particle
($A' = \pi ~ a^{2}$, $a$ is ``effective radius''), $w \equiv w_{1}$,
\begin{equation}\label{5}
w_{j} = \gamma ~ ( 1 ~-~ \vec{v} \cdot \vec{e}_{j} / c ) ~, ~~ j = 1, 2, 3 ~,
\end{equation}
effective factors are given by relations
\begin{eqnarray}\label{6}
Q'_{1} &=& Q'_{ext} ~-~ < \cos \vartheta'> ~ Q'_{sca}  ~,
\nonumber \\
Q'_{2} &=& ~-~ < \sin \vartheta' ~ \cos \varphi ' > ~ Q'_{sca} ~,
\nonumber \\
Q'_{3} &=& ~-~ < \sin \vartheta' ~ \sin \varphi ' > ~ Q'_{sca} ~
\end{eqnarray}
and form components of the vector of radiation pressure efficiency factor;
see references mentioned above. P-R effect immediately
follows from Eq. (1): $Q'_{2} = Q'_{3} =$ 0. Application of general equation
Eq. (1) to Einstein's special case is presented in Kla\v{c}ka (2002b).

\subsection{Equation of motion and its application to Solar System}
Within the accuracy to the first order in $\vec{v} / c$, Eqs. (1) and (4) yield
\begin{eqnarray}\label{7}
\frac{d~ \vec{v}}{d ~t} &=& \frac{S ~A'}{m~c} ~
	      \sum_{j=1}^{3} ~Q_{j} ' ~\left [  \left ( 1~-~ 2~
	      \frac{\vec{v} \cdot \vec{e}_{1}}{c} ~+~
	      \frac{\vec{v} \cdot \vec{e}_{j}}{c} \right ) ~ \vec{e}_{j}
	      ~-~ \frac{\vec{v}}{c} \right ]  ~,
\nonumber   \\
\vec{e}_{j} &=& ( 1 ~-~ \vec{v} \cdot \vec{e'}_{j} / c ) ~ \vec{e'}_{j}  ~+~
	      \vec{v} / c ~, ~~j = 1, 2, 3 ~.
\end{eqnarray}
(We want to stress that values
of $Q'-$coefficients depend on particle's orientation with respect to the
incident radiation -- their values are time dependent.)

In the case of the most simple application to Solar System, gravitational
force of the Sun has to be considered.
Equation of motion of a particle in the gravitational
and electromagnetic radiation fields of the Sun is
\begin{eqnarray}\label{8}
\frac{d~ \vec{v}}{d ~t} &=&  - ~\frac{4~\pi ^{2}}{r^{2}} ~ \vec{e}_{1} ~+~
		\beta ~\frac{4~\pi ^{2}}{r^{2}} ~
		\sum_{j=1}^{3} ~ \frac{Q_{j} '}{Q_{1} '}~ \vec{X}_{j} ~,
\nonumber \\
\vec{X}_{j} &\equiv& \left ( 1~-~ 2~\frac{\vec{v} \cdot \vec{e}_{1}}{c} ~+~
	       \frac{\vec{v} \cdot \vec{e}_{j}}{c} \right )
	     ~ \vec{e}_{j} ~-~ \vec{v} / c  ~,
\nonumber \\
\beta &=& \frac{0.02868}{12 ~ \pi} ~ Q_{1} ' ~
	 \frac{A' \left [ m^{2} \right ]}{m \left [ kg \right ]} ~,
\end{eqnarray}
if length is measured in astronomical unit (AU) and time in years;
$\vec{e}_{1} \equiv \vec{r} / r$.

\section{Fundamental Physics}
One has to take into account relativistic physics when he wants to correctly
understand processes with electromagnetic radiation. Although it may
seem that some electromagnetic phenomena can be completely understood on the
basis of Newtonian physics, it is not true. On the basis of fundamental
work of Planck and Einstein we know that the speed of light in vacuum is $c$
for all observers, concentration of photons depends on the frame of reference.
Thus, it has no sense to say that some astronomically important phenomena
can be explained by violation of the above statements and affirm that
they are based on Newtonian physics (see also Kla\v{c}ka 1992, 1993).

We have relativity theory in disposal for almost one century. We have to
respect its laws if we want to discuss interaction of a particle with
electromagnetic radiation. As a consequence, Eq. (1) is obtained as equation
of motion for the particle under the action of electromagnetic radiation.
Relativistically covariant formulation ensures that ``all inertial observers
are equivalent'', which corresponds to the first postulate of special
relativity. This equivalence of all observers also means that if one inertial
observer carries out some experiments and discovers a physical law, then any
other observer performing the same experiments must discover the same law
(d'Inverno 1992). Applying to our problem, we can affirm: conservations
of energy and momentum in one frame of reference ensures the conservations
also in any other frame of reference.

Relativistic requirement for covariant formulation of equation of motion
leads to Eq. (1) together with three four-vectors
$b_{1}^{\mu}$, $b_{2}^{\mu}$, $b_{3}^{\mu}$ defined by Eq. (4).
Accuracy to the first order in $\vec{v} / c$ reduces Eqs. (1) and (4) to
Eq. (7).

\section{Kimura's access and our result -- comparison}
Kimura's access (Kimura et al. 2002) is based on definitions of two sets
of orthonormal
vectors -- primed and unprimed. On the basis of these definitions
Kimura et al. (2002) obtain their equation of motion.

Eq. (7) shows that orthonormality of the primed unit vectors does not admit
orthonormality of the unprimed unit vectors.

Our access is based on definition of the only one set of orthonormal vectors
-- primed frame of reference (see text below Eq. (4)). We do not need any
another definition! Moreover, relativity principles do not enable
any other definition. Relativity principles strictly determine the expressions
for unprimed unit vectors!

Thus, strict requirement for fulfilling relativity principles leads to
equation of motion which differs from that presented by Kimura et al. (2002).
Since man's knowledge does not doubt about the
requirement for fulfilling relativity principles, Eq. (1) has to be
considered as the right equation of motion. Any other form of equation of motion,
not consistent with Eq. (1), has to be rejected. On this basis one has to
reject Kimura's equation of motion (Eqs. (10), (13) in Kimura et al. 2002).

\section{Physics and Kimura's access}
We have discussed why equation of motion presented in Kimura et al. (2002) 
has to be rejected. We will present physical arguments for the rejection of 
Kimura's equation of motion in a more simple way, now.

Covariant formulation of equation of motion ensures fulfillment of the
principles of relativity theory. As a consequence:
i) if law of conservation of energy is fulfilled in one frame of reference,
then the law conservation of energy is fulfilled in any other frame of reference;
ii) if law of conservation of momentum is fulfilled in one frame of reference,
then the law conservation of momentum is fulfilled in any other frame of reference.

Since equation of motion presented in Kimura et al. (2002) cannot be
formulated in relativistically covariant form, the first principle of
relativity is violated. Thus, an observer in a frame of reference determines
that the laws of conservation of energy and momentum hold during the process
of interaction of the electromagnetic radiation with the particle, while
any other observer not in the same frame of reference will state
that the laws of conservation of energy and momentum do not hold.

Another even more simple argument exists why equation of motion presented in
Kimura et al. (2002) has to be rejected. It is well accepted that a new and more
general theory must contain an older theory as a special case. In other
words, more general theory has to be reducible to the older theory.
As we know, two special cases corresponding to equation of motion for a particle
under the action of electromagnetic interaction, were presented. The first one
was presented by Einstein (1905), the second one by Robertson (1937).
Application of equation of motion presented in Kimura et al. (2002) to the
special case treated by Einstein yields result which is not consistent
with Einstein result -- for more details see Kla\v{c}ka (2002b).

\section{Equation of motion and practical calculations}
We can write for continuous distribution of density flux of energy
as a function of frequency
\begin{equation}\label{9}
\frac{d ~\vec{p'}}{d~ \tau} = \frac{A'}{c} ~
	 \sum_{j=1}^{3} ~\int_{0}^{\infty} ~c~h ~\nu ' ~
	 \frac{\partial n'}{\partial \nu '} ~ Q_{j} '( \nu ') ~ d \nu ' ~
	 \vec{e'}_{j} ~.
\end{equation}
Taking into account that concentration of photons fulfills 
$n' = w ~n$ and $\nu ' = w~ \nu$, we have
$\partial n' ~/~ \partial \nu '$ $=$
$\partial n ~/~ \partial \nu$. Lorentz transformation finally yields
\begin{equation}\label{10}
\frac{d ~p^{\mu}}{d~ \tau} = \frac{w^{2}~A'}{c^{2}} ~
	 \sum_{j=1}^{3} ~ \left (
	 c ~ b_{j}^{\mu} ~-~ u^{\mu}  \right ) ~
	 \int_{0}^{\infty} ~c~h ~
	 \frac{\partial n}{\partial \nu} ~\nu~ Q_{j} '( w~\nu ) ~
	 d \nu ~.
\end{equation}

\subsection{Correct equation of motion}
Instead of incorrect Eq. (13) in Kimura et al. (2002), we
write the right form of equation of motion:
\begin{eqnarray}\label{11}
\frac{d~ \vec{v}}{d ~t} &=& -~ \frac{G~M_{\odot}}{r^{2}} ~ \vec{e}_{1} ~+~
	      \frac{G~M_{\odot}}{r^{2}} ~
	      \sum_{j=1}^{3} ~\beta_{j} ~\left [  \left ( 1~-~ 2~
	      \frac{\vec{v} \cdot \vec{e}_{1}}{c} ~+~
	      \frac{\vec{v} \cdot \vec{e}_{j}}{c} \right ) ~ \vec{e}_{j}
	      ~-~ \frac{\vec{v}}{c} \right ]  ~,
\nonumber   \\
\vec{e}_{j} &=& ( 1 ~-~ \vec{v} \cdot \vec{e'}_{j} / c ) ~ \vec{e'}_{j}  ~+~
	      \vec{v} / c ~, ~~j = 1, 2, 3 ~,
\end{eqnarray}
where
\begin{eqnarray}\label{12}
\beta_{1} &=& \frac{\pi ~R_{\odot}^{2}}{G~M_{\odot}~m~c}
	      \int_{0}^{\infty} ~B_{\odot} ( \lambda ) \left \{
	      C'_{ext} ( \lambda / w ) ~-~ C'_{sca} ( \lambda / w ) ~
	      g'_{1} ( \lambda / w ) \right \} ~ d \lambda ~,
\nonumber   \\
\beta_{2} &=& \frac{\pi ~R_{\odot}^{2}}{G~M_{\odot}~m~c}
	      \int_{0}^{\infty} ~B_{\odot} ( \lambda ) \left \{
	      ~-~ C'_{sca} ( \lambda / w ) ~
	      g'_{2} ( \lambda / w ) \right \} ~ d \lambda ~,
\nonumber   \\
\beta_{3} &=& \frac{\pi ~R_{\odot}^{2}}{G~M_{\odot}~m~c}
	      \int_{0}^{\infty} ~B_{\odot} ( \lambda ) \left \{
	      ~-~ C'_{sca} ( \lambda / w ) ~
	      g'_{3} ( \lambda / w ) \right \} ~ d \lambda ~,
\nonumber   \\
w &=& 1~-~\vec{v} \cdot \vec{e}_{1} ~/~ c ~,
\end{eqnarray}
$R_{\odot}$ denotes the radius of the Sun and $B_{\odot} ( \lambda )$ is
the solar radiance at a wavelength of $\lambda$; $G$, $M_{\odot}$, and
$r$ are the gravitational constant, the mass of the Sun, and the distance
of the particle from the center of the Sun, respectively.
$C'_{ext}$ and $C'_{sca}$ denote the usual extinction and scattering
cross sections, the asymmetry parameter vector $\vec{g}'$ is defined
by $\vec{g}' = \int \vec{n}' ( d C'_{sca} / d \chi ') d \chi '$, where
$d \chi '$ is the element of solid angle, $\vec{n}'$ is a unit vector
in the direction of scattering, $d C'_{sca} / d \chi '$ is the differential
scattering cross section; cartesian coordinates $( g'_{1}, g'_{2}, g'_{3} )$
describe the asymmetry parameter vector $\vec{g}'$ as
$\vec{g}'$ $=$ $g'_{1} ~ \vec{e}'_{1}$ $+$ $g'_{2} ~ \vec{e}'_{2}$ $+$
$g'_{3} ~ \vec{e}'_{3}$ -- see Kimura et al. (2002), as for comparison see
Kla\v{c}ka and Kocifaj (2001a).

Eq. (11) corresponds to Eq. (8): $\beta_{j} = \beta~ Q'_{j}~/~Q'_{1}$,
$j =$ 1, 2, 3. Eq. (12) is equivalent to Eq. (23) in Kla\v{c}ka and
Kocifaj (2001a). As a consequence of the fact that unit vectors
$\vec{e}_{1}$, $\vec{e}_{2}$, $\vec{e}_{3}$ do not form orthonormal set, 
we may mention that equation
$\vec{v}$ $=$ $( \vec{v} \cdot \vec{e}_{1} ) \vec{e}_{1}$ $+$
$( \vec{v} \cdot \vec{e}_{2} ) \vec{e}_{2}$
is not fulfilled.

\section{Conclusion}
Equation of motion presented by Kimura et al. (2002 -- Eqs. (10), (13)) 
does not respect basic principles of relativity. Kimura et al. (2002) 
use physically unacceptable definitions. As a consequence,
application of equation of motion presented by Kimura et al. (2002) to the
special case treated by Einstein yields result which is not consistent
with Einstein result. All these facts lead to the conclusion:
equation of motion presented in Kimura et al. (2002) has to be rejected.

Correct equation of motion is presented in Kla\v{c}ka (2000a, 2000b, 2000c,
2001, 2002a), Kla\v{c}ka and Kocifaj (2001a) and in some other later
papers of these authors -- all these papers present and use
equation of motion consistent with Eq. (1) of this paper. If one wants to
use correct equation of motion which contains the quantities used by
Kimura et al. (2002), Eqs. (11) and (12) of our paper are correct.

\acknowledgements{This paper was supported by the Scientific Grant Agency VEGA
grant No. 1/7067/20.}


\begin{thebibliography}{}
\bibitem{}d'Inverno R. 1992. Introducing Einstein's Relativity.
Oxford University Press, Oxford, pp. 383.
\bibitem{}Einstein A. 1905. Zur Elektrodynamik der bewegter K\H{o}rper.
{\it Annalen der Physik} {\bf 17}, 891-920.
\bibitem{}Kimura H. 2000. Light-scattering properties of fractal
aggregates: numerical calculations by superposition technique and
discrete-dipole approximation.
In: {\it Light Scattering by Nonspherical Particles: Halifax Contributions},
G. Videen, Q. Fu, and P. Ch\'{y}lek (eds.), Army Research Laboratory,
Adelphi Maryland, pp. 241-244.
\bibitem{}Kimura H., Okamoto H., Mukai T. 2002.
Radiation Pressure and the Poynting-Robertson Effect
for Fluffy Dust Particles. {\it Icarus} {\bf 157}, 349-361.
\bibitem{}Kla\v{c}ka J. 1992. Poynting-Robertson effect. I. Equation of motion.
{\it Earth, Moon, and Planets} {\bf 59}, 41-59.
\bibitem{}Kla\v{c}ka J. 1993. Misunderstanding of the Poynting-Robertson effect.
{\it Earth, Moon, and Planets} {\bf 63}, 255-258.
\bibitem{}Kla\v{c}ka J. 2000a. Electromagnetic radiation and motion of real particle. \\
http://xxx.lanl.gov/abs/astro-ph/0008510 (Icarus: manuscript 7883,
rejected in 2000)
\bibitem{}Kla\v{c}ka J. 2000b. Aberration of light and motion of real particle. \\
http://xxx.lanl.gov/abs/astro-ph/0009108
\bibitem{}Kla\v{c}ka J., 2000c, Solar radiation and asteroidal motion. \\
http://xxx.lanl.gov/abs/astro-ph/0009109
\bibitem{}Kla\v{c}ka J. 2001. Motion of electrically neutral particle in the
field of electromagnetic radiation.
{\it Meteor Reports} {\bf 22}, 21-28.
\bibitem{}Kla\v{c}ka J. 2002a. Covariant equation of motion for a particle in
an electromagnetic field. In: {\it Optics of Cosmic Dust}, G. Videen and
M. Kocifaj (eds.), Kluwer Academic Publishers, Dordrecht/Boston/London,
pp. 301-312.
\bibitem{}Kla\v{c}ka J. 2002b. On equation of motion for arbitrarily shaped
particle under action of electromagnetic radiation. 
http://xxx.lanl.gov/abs/astro-ph/0204148
%\bibitem{}Kla\v{c}ka J., 2002b, Orbital elements for motion of real particle
%under the action of electromagnetic radiation. \\
%http://xxx.lanl.gov/abs/astro-ph/0201201
\bibitem{}Kla\v{c}ka J., Kocifaj M. 2001a. Motion of nonspherical dust particle
under the action of electromagnetic radiation.
{\it J. Quant. Spectrosc. Radiat. Transfer} {\bf 70/4-6}, 595-610.
\bibitem{}Kla\v{c}ka J., Kocifaj M. 2001b. On the stability of the zodiacal
cloud. In: Dynamics of Natural and Artificial Celestial Bodies,
H. Pretka-Ziomek, E. Wnuk, P. K. Seidelmann and D. Richardson
(eds.), Kluwer Academic Publishers, Dordrecht, pp. 355-357.
\bibitem{}Kla\v{c}ka J., Kocifaj M. 2002a. Temporary capture of dust grains in
exterior resonances with Earth.
In: {\it Electromagnetic and Light Scattering by Nonspherical Particles},
B. A. S. Gustafson, L. Kolokolova and G. Videen (eds.), Army Research Laboratory,
Adelphi Maryland, pp. 167-169.
\bibitem{}Kocifaj M., Kla\v{c}ka J. 2002a. On the spread of a micron-sized
fraction of the dust grain population from comet Encke.
In: {\it Electromagnetic and Light Scattering by Nonspherical Particles},
B. A. S. Gustafson, L. Kolokolova and G. Videen (eds.), Army Research Laboratory,
Adelphi Maryland, pp. 171-174.
\bibitem{}Kocifaj M., Kla\v{c}ka J. 2002b. The capture of interstellar dust.
The pure electromagnetic radiation case.
{\it Planet. Space Sci.} (submitted).
\bibitem{}Kocifaj M., Kla\v{c}ka J., Kundrac\'{\i}k F. 2000.
Motion of realistically shaped cosmic dust particle in Solar System.
In: {\it Light Scattering by Nonspherical Particles: Halifax Contributions},
G. Videen, Q. Fu, and P. Ch\'{y}lek (eds.), Army Research Laboratory,
Adelphi Maryland, pp. 257-261.
\bibitem{}Kla\v{c}ka J., Kocifaj M. 1994. Electromagnetic radiation and
equation of motion for a dust particle.
In: Dynamics and Astrometry of Natural and
Artificial Celestial Bodies, K. Kurzy\'{n}ska, F. Barlier, P. K. Seidelmann
and I. Wytrzyszczak (eds.), Astronomical Observatory of A. Mickiewicz
University, Pozna\'{n}, Poland, 187-190.
%\bibitem{}Purcell E. M. 1979. Suprathermal rotation of interstellar grains.
%{\it Astrophys. J.} {\bf 231}, 404-416.
\bibitem{}Robertson H. P. 1937. The dynamical effects of radiation in the Solar
System. {\it Mon. Not. R. Astron. Soc.} {\bf 97}, 423-438.
\end{thebibliography}
\end{document}